\documentclass[letterpaper, 10 pt, conference]{ieeeconf}  

\IEEEoverridecommandlockouts                              

\usepackage{mathptmx} 
\usepackage{times} 
\usepackage{amssymb}  
\usepackage{graphicx}
\usepackage{amsmath, nccmath}
\usepackage{tabularx,ragged2e,booktabs,caption}
\usepackage{geometry}
\usepackage{cite}

\title{\LARGE \bf
Learning-based Model Predictive Control for Smart Building Thermal Management }

\author{Roja Eini$^{1}$ and Sherif Abdelwahed$^{2}$
    \thanks{$^{1}$Roja Eini is PhD student of Department of Electrical Engineering,
        Virginia Commonwealth University, Richmond, VA, USA
        {\tt\small einir@vcu.edu}}%
    \thanks{$^{2}$Sherif Abdelwahed is faculty of Department of Electrical Engineering,
        Virginia Commonwealth University, Richmond, VA, USA
        {\tt\small sabdelwahed@vcu.edu}}%
}

\begin{document}

\maketitle
\thispagestyle{empty}
\pagestyle{empty}

\begin{abstract}
This paper proposes a learning-based model predictive control (MPC) approach for the thermal control of a four-zone smart building. The objectives are to minimize energy consumption and maintain the residents' comfort. The proposed control scheme incorporates learning with the model-based control. The occupancy profile in the building zones are estimated in a long-term horizon through the artificial neural network (ANN), and this data is fed into the model-based predictor to get the indoor temperature predictions. The Energy Plus software is utilized as the actual dataset provider (weather data, indoor temperature, energy consumption). The optimization problem, including the actual and predicted data, is solved in each step of the simulation and the input setpoint temperature for the heating/cooling system, is generated. Comparing the results of the proposed approach with the conventional MPC results proved the significantly better performance of the proposed method in energy savings ($40.56 \%$ less cooling power consumption and $16.73 \%$ less heating power consumption), and residents' comfort.  
\end{abstract}  

\section{INTRODUCTION}
The residential and commercial building sector is known to use around 40\% of the total end-use energy and, hence, is considered to be the largest energy-consuming sector in the world [1]. Approximately half of this energy is used for heating, cooling, ventilation, and air conditioning (HVAC), and this usage is increasing $0.5-5$\% per year in developed countries [2]. This trend is similar to the rest of the world. Therefore, finding solutions to reduce energy use and/or increase energy efficiency in the building sectors, particularly for smart buildings in the smart city environment, is of crucial importance.

The majority of building thermal controls are based on model-based approaches. In model-based control designs, the controller is designed based on the mathematical model of the plant, assuming that the model represents the actual plant. However, model uncertainties and modeling errors always exist in the modeling process. 
One of the efficient model-based techniques in building thermal control is model predictive control (MPC) [3]. Just like other model-based strategies, MPC requires an accurate model (mathematical model) to predict the process inputs/outputs and obtain the control signal [4-6]. The performance of MPC is directly relevant to the accuracy of the model, and it diminishes by the model inaccuracy. In the context of building management, it is difficult to accurately identify the building's thermal models, due to vast differences in construction materials and architectures, time-varying thermal dynamics, huge load of complex data processing, high cost of accurate modeling, and difficulties in modeling the residents' behavior and occupancy.

Learning-based modeling is known to be efficient in accurate modeling of multi-zone buildings with nonlinearities, uncertainties, time-varying characteristics, the vast number of variables and components, nonuniform zone temperatures, and zonal couplings [7]. A machine learning algorithm can utilize the building's historical datasets to improve the modeling or control framework over time by learning the model uncertainties and real-life conditions [8]. Machine learning algorithms can address the complex data such as occupant behavior and varying operating costs in a building management system, without requiring a detailed model and explicit programming [9].

By integrating learning-based algorithms with model-based controls, one can utilize both advantages of learning- and modeling-based designs. On the one hand, model-based design assists the learning-based design in learning explorations with maximum learning rate. On the other hand, occupants' behaviors, and various cyber physical interactions can be handled by the learners and fed into the model-based management structure. The use of learning-based algorithms in building modeling and control have been studied in the literature [10]-[12]. In [10], an ANN model is used to decrease the temperature overshoot and undershoots in the HVACs, which result in the reduction of energy consumption. Authors in [11] and [12] employed feed-forward ANN to build a predictive thermal model to determine the ON/OFF time for the HVAC. None of the mentioned works considered the occupancy profile in the learning process. Moreover, none of them incorporated a model-based control framework with the learning-based algorithm to get the most advantage of both designs.

In this paper, a learning-based modeling strategy is incorporated with a model-based predictive control algorithm to manage a multi-zone building's thermal property. An ANN is utilized to predict the occupancy profile; then this data is fed into the model-based control (MPC). The datasets of ANN are generated by simulating an actual building in Energy Plus software, considering the indoor temperature, time of day, weather data, energy consumption data, and setpoint temperatures. Through the MPC algorithm, the optimum setpoints are generated as the control inputs at each step to conserve energy and improve the comfort level. In contrast to the previous works, this work utilized both model-based and learning-based modeling in the MPC algorithm to enhance robustness and stability of the model-based control framework as well as improving the controlled system performance through learning from historical datasets.  \\ \indent
The rest of the paper is organized as follows. Section II describes the building model and its components. Section III and IV explain the learning-based and model-based modeling techniques, respectively. The next section introduces the proposed learning-based MPC approach. The simulation results and simulation assumptions are presented in section VI. Finally, section VII provides conclusions and discusses future research.   

\section{SYSTEM DEFINITION}
\begin{figure}[h]
    \begin{center}
                \centering
                \includegraphics[height=6cm, width=8.5cm]{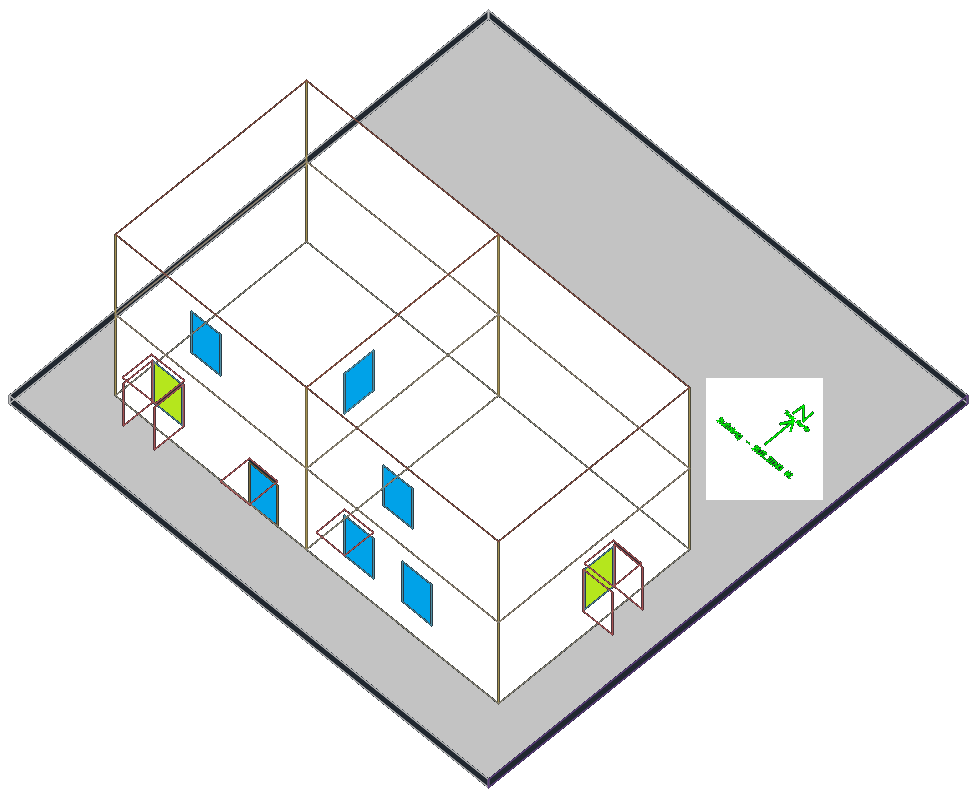}
                \caption{four-zone building CAD model}
        \label{fig:building}
    \end{center}
\end{figure}

A two-story office building with four zones and one HVAC system per zone is considered in this work. Each zone thermostat is dual setpoint.  Fig. \ref{fig:building} shows the four-zone building CAD model. The floor area is $1600 m^2$ with the orientation to the north. Windows include shadings, overhangs, and fins. Several materials are used in various layers of the walls (exterior and interior), window frames, door, roof, ceiling, and inter-zone walls. Table \ref{tab:systemparams} contains the building materials' specifications.  

\begin{table*}[ht]
        \caption{Building materials description}
        \begin{tabular}{| c | m{1.3cm} | m{1.3cm} | m{1.3cm} | m{1.3cm} | m{1.3cm} | m{1.3cm} | m{1.3cm} | m{1.3cm} | m{1.3cm} |} 
            \hline
            
            & 4 inch dense face brick & {2 inch insulation} & {4 inch concrete block} & {3/4 inch plaster board} & {1/8 inch hardwood} & {8 inch concrete block} & {acoustic tile} & {1/2 inch stone} & {3/8 inch membrane} \\  \hline
            {Roughness} & {Rough} & {Very rough} & {Medium rough} & {Smooth} & {Medium smooth} & {Rough} & {Medium smooth} & {Rough} & {Rough}\\ \hline
            {Thickness \tiny $(m)$} & $0.1014684$ & $0.050901$ & $0.1014984$ & $0.019050$ & $0.003169$ & $0.2033016$ & $0.019050$ & $0.012710$ & $0.009540$\\ \hline
            {Conductivity \tiny $(W/m-K)$} & $1.245296$ & $0.043239$ & $0.3805070$ & $0.7264224$ & $0.1591211$ & $0.5707605$ & $0.060535$ & $1.435549$ & $0.1902535$\\ \hline
            {Density \tiny $(kg/m^3)$} & $2082.400$ & $32.03693$ & $608.7016$ & $1601.846$ & $720.8308$ & $608.7016$ & $480.5539$ & $881.0155$ & $1121.292$\\ \hline
            {Specific heat \tiny $(J/kg-K)$} & $920.4800$ & $836.8000$ & $836.8000$ & $836.8000$ & $1255.200$ & $836.8000$ & $836.8000$ & $1673.600$ & $1673.600$\\ \hline
            {Thermal absorptance} & $0.900000$ & $0.900000$ & $0.900000$ & $0.900000$ & $0.900000$ & $0.900000$ & $0.900000$ & $0.900000$ & $0.900000$\\ \hline
            {solar absorptance} & $0.930000$ & $0.500000$ & $0.650000$ & $0.920000$ & $0.780000$ & $0.650000$ & $0.320000$ & $0.550000$ & $0.750000$\\ \hline
            {Visible absorptance} & $0.930000$ & $0.500000$ & $0.650000$ & $0.920000$ & $0.780000$ & $0.650000$ & $0.320000$ & $0.550000$ & $0.750000$\\ \hline
        \end{tabular}
        \label{tab:systemparams}
\end{table*}

\section{OCCUPANCY PREDICTIONS}
\begin{figure}[t]
    \begin{center}
                \centering
                \includegraphics[height=3.3cm, width=6.5cm]{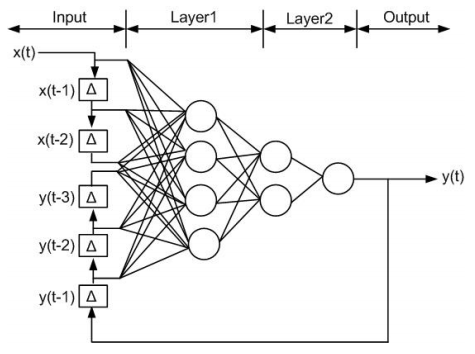}
        \caption{A typical NARX network with input and output delays}
        \label{fig:ANN_typical}
    \end{center}
\end{figure}

This section explains the learning-based approach to predict the occupancy impact on the indoor temperature. Occupancy information is important in energy-efficient building climate control since it can impact the temperature, environmental conditions, energy usage, and comfort constraints [2]. The goal of this work is to predict the occupancy information in the long-term through the ANN and investigate its influence on the building energy consumption and residents' comfort. The selected ANN model is a Nonlinear Autoregressive netwoRk with eXogenous inputs (NARX). In the NARX model, at least three layers of nodes (input, output, and hidden layer) are used to approximate the outputs in (\ref{eq:NARX}). \\
\begin{fleqn}
    \begin{equation} 
    \begin{aligned}[b] 
    y(t)&=f(x(t-1),...,x(t-d_x),y(t-1),y(t-2),...,y(t-d_y)) 
    \end{aligned}
    \label{eq:NARX}
    \end{equation}
\end{fleqn} 
$x(t)$, $y(t)$, $d_x$, and $d_y$ denote the inputs, outputs, input delays, and output delays of the ANN, respectively. $f$ is the mapping function. There are generally two architectures of NARX neural networks; i.e., series-parallel architecture and parallel architecture. 
The series-parallel configuration provides better prediction performance in training the model than the parallel configuration. The better performance is because the input of the feed-forward network is more accurate, and the static backpropagation can be used for training. Since in our work the actual output is available during the training of the network (from Energy Plus data), we have chosen the series-parallel architecture to train the network. The series-parallel NARX network is represented in Fig. \ref{fig:ANN_typical}.

The reason we used the NARX neural network for the occupancy predictions is that the occupancy profile is a time series, and one of the primary applications of NARX is predicting the time series models [13]. Moreover, since the occupancy produces heating, it is highly nonlinear, and NARX model is very beneficial for nonlinear models of this type. 
After training the network, it is validated. To evaluate the stopping criterion and the expected performance of the predicted data, the test data is used. Therefore, three datasets are used; training, validation, and test.  Mean square error (MSE) and the regression value, representing the square error and the correlation between the output and the target values are utilized to validate the training performance.
Thus, the NARX neural network algorithm is as follows:

\begin{itemize}
\item Define the input and output datasets. 
\item Define three sets of training, validation, and testing data.
\item Choose a network architecture and a  training algorithm by trial and error method. 
\item Train the network, and evaluate its performance.  
\item If the network performance is satisfactory, the problem is solved, otherwise, change the network size, retrain, or use larger datasets. 
\end{itemize}

The ANN specifications of this work are described in section VI.
 
\section{INDOOR TEMPERATURE PREDICTIONS}
This section explains the model-based approach to predict the indoor temperature. Considering the thermal convection and conduction equations, the mathematical model of the indoor temperature is represented as (\ref{eq:thermal})  [14], [15]. \\
\begin{fleqn}
    \begin{equation} 
    \begin{aligned}[b] 
    \hat T_{in}(t)&=a[\hat T_{in}(t-1)+\frac {\Delta t}{C}[P(t-1)-U(\hat T_{in}(t-1)- T_{out}(t-1))]]  \\
    &+\hat b(t)
    \end{aligned}
    \label{eq:thermal}
    \end{equation}
\end{fleqn} 
\noindent
where $\hat T_{in}$ and $T_{out}$ are the estimated indoor temperature and the outdoor temperature, respectively. $\Delta t$ is the time step, and $P$ is the heating power. $a$ and $U$ are the parameters to be identified. $\hat b(t)$ is the estimated occupancy at time $t$. In the learning-based simulation, the estimated value of occupancy is fed into the model-based predictor. In the conventional MPC; i.e., without learning, the occupancy profile is chosen constant at its average value ($\bar b(t)$). \\ \indent
The parameters of the thermal model (\ref{eq:thermal}) are identified through the recursive least square (RLS) identification algorithm using the Energy Plus input/output data. To evaluate the identification algorithm performance, the root mean square (RMS) criterion is used. The RLS algorithm is represented in brief as follows. \\
\begin{fleqn}
    \begin{equation} 
    \begin{aligned}[b] 
    &\hat F(t+1)=\frac{1}{\lambda}[F(t)-\frac{F(t) {\phi}^T(t) F(t)}{\lambda +{\phi}^T(t)F(t)\phi (t)}] \\
    &e(t+1)=y(t+1)-\hat \theta (t) \phi (t)  \\
    &\hat \theta (t+1)=\hat \theta (t) + F(t+1) \phi (t) e(t+1)
    \end{aligned}
    \label{eq:RLS}
    \end{equation}
\end{fleqn}
\noindent
where $F$, $\lambda$, $\phi$, and $\hat \theta$ are the gain, forgetting factor, observations and estimated parameter, respectively. $e$ represents the error between the measurements and identification outputs.  

\section{LEARNING-BASED MODEL PREDICTIVE CONTROL}
\begin{figure}[thb]
    \begin{center}
                \centering
                \includegraphics[height=6.5cm, width=9cm]{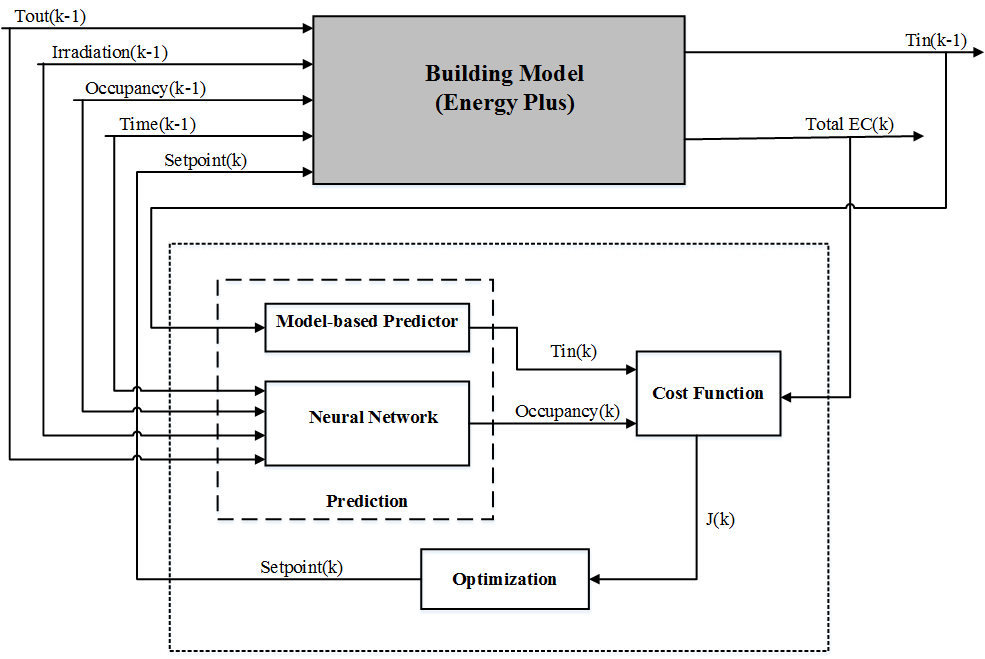}
        \caption{Learning-based model predictive control (MPC)}
        \label{fig:block}
    \end{center}
\end{figure}
Having the weather and occupancy forecasts, the model predictive control (MPC) comes into play. MPC is a modern control technique that has been applied in many areas due to its ability to handle constrained control problems [3]. At each time instant, an optimal control problem is solved to obtain the optimal control action over the time horizon. Using MPC in the building temperature control, a plan for the HVAC system control is generated based on the predicted weather conditions and occupancy profiles over the time horizon. The first control action that minimizes the energy consumption and satisfies the comfort is applied to the building's HVACs, then the control algorithm is repeated with the feedback information of building states and outputs at the next time instant. Fig. \ref{fig:block} represents the proposed learning-based model predictive control approach. \\ \indent
MPC cost function is defined as (\ref{eq:cost}), such that it penalizes the deviations from the comfort level and optimum energy consumption. \\
\begin{fleqn}
    \begin{equation} 
    \begin{aligned}[b] 
    &J(t)=\sum\limits_{k=0}^{N} {\lVert{\hat T_{in}(t+k) - T_d\lVert}_Q }^2 + \sum\limits_{k=0}^{N} {\lVert{\Delta P(t+k-1)\lVert}_R }^2
    \end{aligned}
    \label{eq:cost}
    \end{equation}
\end{fleqn}
 \noindent
where $Q$ and $R$ are the weighting factors associated with the states and inputs, respectively. $N$ is the time horizon, and $T_d$ is the comfort setpoint temperature. Therefore, the MPC problem is to minimize (\ref{eq:cost}) subject to the performance constraints (\ref{eq:constraints1}), robustness constraints (\ref{eq:constraints2}), and limit constraints ((\ref{eq:constraints3})). It is worth mentioning that equation (\ref{eq:constraints1}) incorporates the learning while (\ref{eq:constraints2}) is solely based on model-based design. 
\begin{fleqn}
    \begin{equation} 
    \begin{aligned}[b] 
    \hat T_{in}(t)&=a[\hat T_{in}(t-1)+\frac {\Delta t}{C}[P(t-1)-U(\hat T_{in}(t-1)- T_{out}(t-1))]]  \\
    &+\hat b(t)
    \end{aligned}
    \label{eq:constraints1}
    \end{equation}
\end{fleqn} 
\begin{fleqn}
    \begin{equation} 
    \begin{aligned}[b] 
    \bar T_{in}(t)&=a[\bar T_{in}(t-1)+\frac {\Delta t}{C}[P(t-1)-U(\bar T_{in}(t-1)- T_{out}(t-1))]]  \\
    &+\bar b(t)
    \end{aligned}
    \label{eq:constraints2}
    \end{equation}
\end{fleqn}

\begin{fleqn}
    \begin{equation} 
    \begin{aligned}[b] 
    & T_{in}^{min} \leq \bar T_{in}(t+k) \leq T_{in}^{max},  \\
    & P^{min} \leq P(t+k-1) \leq P^{max}
    \end{aligned}
    \label{eq:constraints3}
    \end{equation}
\end{fleqn}\\
MPC algorithm is as follows:  
\begin{itemize}
\item Define the system states and inputs at the current time, and their estimations up to the time horizon. 
\item Solve the optimization problem (cost function) to get the optimum inputs at time $t$.
\item At time $t$, solve the optimization to get the input signal over the horizon.
\item Apply the first control input, $t=t+1$, and go to the second step. 
\end{itemize}

\section{SIMULATION RESULTS}
In this section, all the simulation assumptions and results from the proposed learning-based MPC and the conventional MPC (without learning) are illustrated. The simulations are performed for one year, with 6 time steps per hour. To provide the ANN dataset, Energy Plus simulations on the building model of Fig. \ref{fig:building} were completed from the 1st of July to 31st of December. The simulation assumptions are as follows. \\
\begin{figure}[t]
\center
                \includegraphics[height=2.5cm, width=0.48\textwidth]{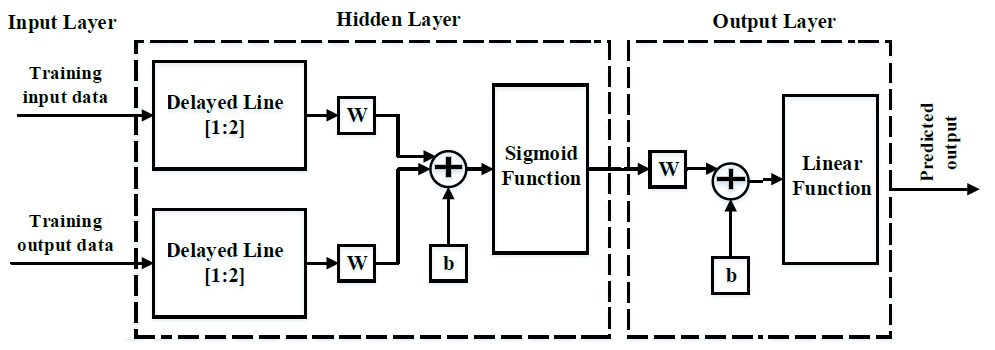}
                \caption{NARX neural network series model}
        \label{fig:NN}
\end{figure}
\begin{figure}[t]
    \begin{center}
                \centering
                \includegraphics[height=5cm, width=0.48\textwidth]{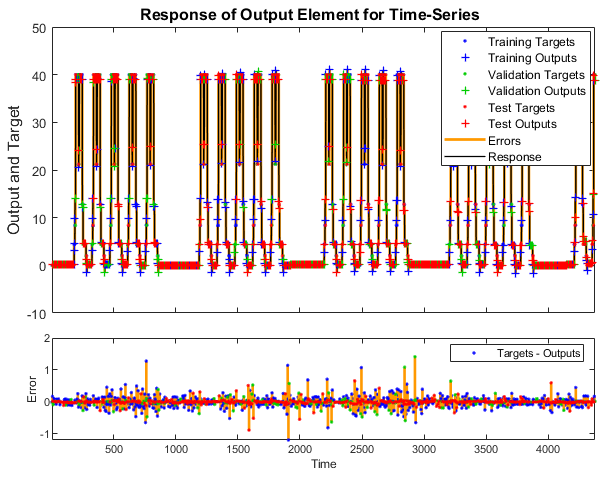}
        \caption{Neural network output response versus targets}
        \label{fig:timeseries}
    \end{center}
\end{figure}
\begin{itemize}
    \item The desired temperature of all zones are between 20 $^\circ$C and 25 $^\circ$C.
    \item The control variables are the HVAC setpoints.
    \item The maximum and minimum supply air temperatures are 50 $^\circ$C and 13 $^\circ$C, respectively.
    \item The maximum dry-bulb temperature for winter and summer days in Chicago Ohare location are considered -16.6 $^\circ$C and 31.6 $^\circ$C, respectively.
    \item The weather data at Chicago Ohare location is used. 
    \item The number of people per zonal area is 0.1.
    \item The ANN input layer includes the environmental measures; the time of day, date, weather data, and the historical occupancy data.
    \item The input and output delays of NARX model are both chosen 2. 
    \item One output layer and 10 hidden layers are chosen. 
    \item The Levenbegrg-Marquardt backpropagation training algorithm is chosen. 
\end{itemize} 
\begin{figure}[t]
    \begin{center}
                \centering
                \includegraphics[height=4.9cm, width=0.5\textwidth]{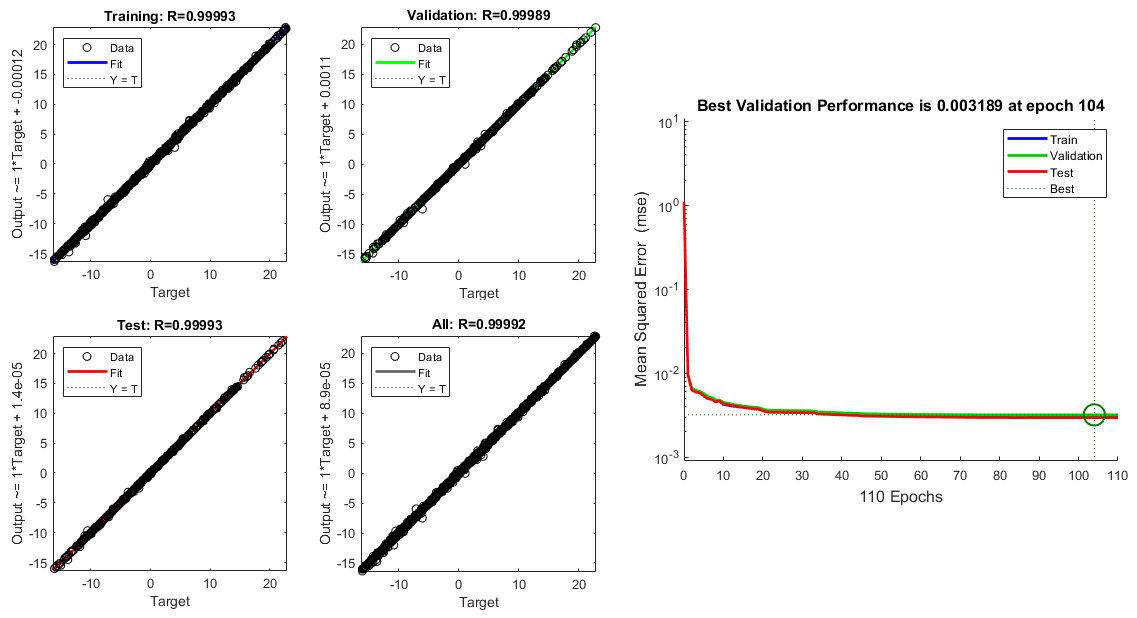}
        \caption{Regression and performance trajectories of datasets}
        \label{fig:performance}
    \end{center}
\end{figure}
\begin{figure}[t]
    \begin{center}
                \centering
                \includegraphics[height=4.8 cm, width=0.4\textwidth]{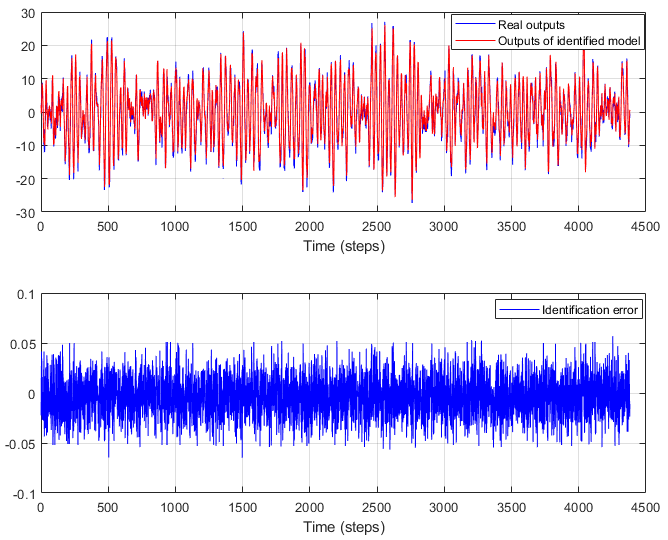}
        \caption{The identified model outputs versus real outputs, and the identification error trajectory}
        \label{fig:id}
    \end{center}
\end{figure}

\indent
The NARX neural network implemented in MATLAB is presented in Fig. \ref{fig:NN}. Fig. \ref{fig:timeseries} compares the network's response with the actual vacancy profile, and shows the error values between the occupancy predictions and its actual profile throughout one month (To get a clear image, these plots are presented for one-month period). The maximum error value at each time is $1$; i.e., the target occupancy profile is well-tracked. Fig. \ref{fig:performance} presents the regression and performance plots of the training, validation, and testing datasets. The regression values are all close to $1$ and the MSE error is $0.003189$; i.e., the training performance is satisfactory.
\begin{figure}[t]
    \begin{center}
                \centering
                \includegraphics[width=0.48\textwidth]{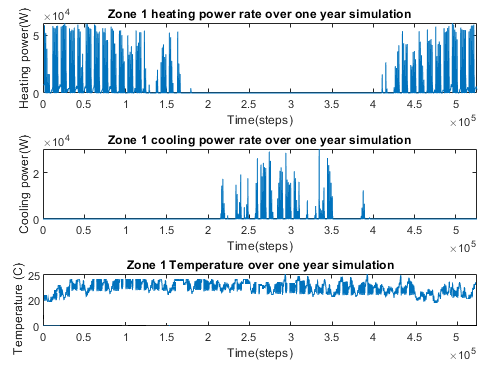}
        \caption{The heating/cooling power consumption rate and zone 1 temperature using the proposed learning-based MPC}
        \label{fig:proposed} 
            \end{center}
\end{figure}  

\begin{figure}[t]
    \begin{center}
                \centering
                \includegraphics[width=0.48\textwidth]{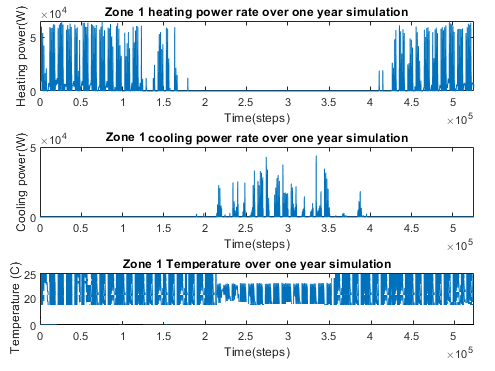}
        \caption{The heating/cooling power consumption rate and zone 1 temperature using the conventional MPC}
        \label{fig:conv}
    \end{center}
\end{figure}

\begin {table} [t]
\caption {Simulation results}
\begin{center}
    \noindent
    \newlength{\myheight}
    \setlength{\myheight}{0.5cm}
    \footnotesize
    \begin{tabular}{| c | m{1.4cm} | m{1.4cm} | m{1.1cm} |}  
        \hline
        \parbox[c][\myheight][c]{0cm}{}   Parameters \small &  Conventional MPC  \small & Learning-based MPC  \small &  Change  \small \\  
        \hline \hline 
        \parbox[c][\myheight][c]{0cm}{}    Average cooling power  \small & 396.28 W  \small & 235.55 W  \small & $\downarrow$ 40.56\%  \small  
        \\   
        \parbox[c][\myheight][c]{0cm}{}    Average heating power \small & 2.43 KW \small & 2.02 KW \small &  $\downarrow$ 16.73\% \small
        \\ 
        \hline \hline 
    \end{tabular}
    \label{tab:Simulation results} 
\end{center}
\end {table}  
Fig. \ref{fig:id} shows the results of indoor temperature model identification throughout one-month simulation. From Fig. \ref{fig:id}, the identification error does not exceed 0.05; i.e., the identified outputs (indoor temperature) are very close to the actual indoor temperature values. Figs. \ref{fig:proposed} and \ref{fig:conv} show the results of the proposed learning-based MPC and conventional MPC approaches on the building throughout one-year simulation. Comparing the power rate graphs and Table \ref {tab:Simulation results} values using the two approaches, the proposed method decreased the cooling and heating power consumption by $40.56\%$ and $16.73 \%$, respectively. Furthermore, the deviations from the comfort level in the conventional approach is extremely higher compared to the proposed method. The zone temperature using the conventional MPC even violates the minimum comfort level. 

\section{CONCLUSIONS}
In this paper, a learning-based MPC strategy is introduced to control the thermal property of a four-zone office building. Predicting the building parameters is a crucial and challenging part of MPC since the building's thermal model is nonlinear, associated with uncertainties, and strongly coupled. Thus, ANN is incorporated with the model- based control approach to address the mentioned issues. The occupancy profile predictions are generated through ANN, and then this data is fed into the model-based controller. Energy Plus software is used in this work to simulate a building with real materials and components, and to test the proposed approach on it. Results from the proposed learning-based approach showed significantly better performance, in maintaining the residents' comfort and minimizing energy usage ($40.56 \%$ energy savings), compared to the conventional MPC. For future work, implementing learning-based control to consider the impact of occupants behavior, such as window opening, or the energy storage devices in the building management system will be considered.     


\begin{thebibliography} {99}
\bibitem{c1} Pérez-Lombard, Luis, José Ortiz, and Christine Pout. ``A review on buildings energy consumption information." Energy and buildings 40, no. 3 (2008): 394-398.
\bibitem{c2} Gul, Mehreen S., and Sandhya Patidar. ``Understanding the energy consumption and occupancy of a multi-purpose academic building." Energy and Buildings 87 (2015): 155-165.
\bibitem{c3} Serale, Gianluca, Massimo Fiorentini, Alfonso Capozzoli, Daniele Bernardini, and Alberto Bemporad. ``Model predictive control (MPC) for enhancing building and HVAC system energy efficiency: Problem formulation, applications and opportunities." Energies 11, no. 3 (2018): 631.
\bibitem{c4} Zohrabi, Nasibeh, Sherif Abdelwahed, and Jian Shi. ``Reconfiguration of MVDC shipboard power systems: A model predictive control approach." In 2017 IEEE Electric Ship Technologies Symposium (ESTS), pp. 253-258. IEEE, 2017.
\bibitem{c5} Eini, Roja, and Sherif Abdelwahed. ``Urban Traffic Network Control in Smart Cities; a Distributed Model-based Control Approach." arXiv preprint arXiv:1905.09955 (2019).
\bibitem{c6} Morrissett, Adam, Roja Eini, Mostafa Zaman, Nasibeh Zohrabi, and Sherif Abdelwahed. ``A Physical Testbed for Intelligent Transportation Systems." arXiv preprint arXiv:1907.12899 (2019).
\bibitem{c7} Klein J., ``Machine learning perspectives for smart buildings: an overview," 2017.
\bibitem{c8} Pasandi, Hannaneh Barahouei, and Tamer Nadeem. ``Challenges and Limitations in Automating the Design of MAC Protocols Using Machine-Learning." In 2019 International Conference on Artificial Intelligence in Information and Communication (ICAIIC), pp. 107-112. IEEE, 2019.
\bibitem{c9} Barahouei Pasandi, Hannaneh, and Tamer Nadeem. ``Poster: Towards Self-Managing and Self-Adaptive Framework for Automating MAC Protocol Design in Wireless Networks." In Proceedings of the 20th International Workshop on Mobile Computing Systems and Applications, pp. 171-171. ACM, 2019.
\bibitem{c10} Moon, Jin Woo, and Jong-Jin Kim. ``ANN-based thermal control models for residential buildings." Building and Environment 45, no. 7 (2010): 1612-1625.
\bibitem{c11} Ruano, Antonio E., Eduardo M. Crispim, Eusébio ZE Conceiçao, and Ma Manuela JR Lúcio. ``Prediction of building's temperature using neural networks models." Energy and Buildings 38, no. 6 (2006): 682-694.
\bibitem{c12} Lu, Tao, and Martti Viljanen. ``Prediction of indoor temperature and relative humidity using neural network models: model comparison." Neural Computing and Applications 18, no. 4 (2009): 345.
\bibitem{c13} Diaconescu, Eugen. ``The use of NARX neural networks to predict chaotic time series." Wseas Transactions on computer research 3, no. 3 (2008): 182-191.
\bibitem{c14} Eini, Roja, and Sherif Abdelwahed. ``Distributed Model Predictive Control Based on Goal Coordination for Multi-Zone Building Temperature." In 2019 IEEE Green Technologies Conference (GreenTech), Lafayette, LA. 2019.
\bibitem{c15} Eini, Roja, Lauren Linkous, Nasibeh Zohrabi, and Sherif Abdelwahed. ``A testbed for a smart building: design and implementation." In Proceedings of the Fourth Workshop on International Science of Smart City Operations and Platforms Engineering, pp. 1-6. ACM, 2019.

\end{thebibliography}
\end{document}